\documentclass{article}

\usepackage{PRIMEarxiv}

\usepackage[utf8]{inputenc} 
\usepackage[T1]{fontenc}    
\usepackage{hyperref}       
\usepackage{url}            
\usepackage{booktabs}       
\usepackage{tabularx}
\usepackage{amsfonts}       
\usepackage{nicefrac}       
\usepackage{microtype}      
\usepackage{lipsum}
\usepackage{fancyhdr}       
\usepackage{graphicx}       
\graphicspath{{media/}}     

\title{PAIRED: A Process-Anchored Framework for Transparent Reporting of AI Contributions in Scientific Research
}

\author{
  Ahmad Al-Kabbany \\
  Multimedia Interaction and Communication Lab \\
  Arab Academy for Science and Technology, Alexandria 21937, Egypt \\
  \texttt{alkabbany@ieee.org}
}

\begin{document}
\maketitle

\begin{abstract}
The rapid integration of generative AI into scientific research has exposed a critical gap in academic disclosure practice. Existing frameworks for reporting AI contributions are uniformly output-oriented — they document what AI produced, not how the research unfolded. As a result, researchers who wish to report their AI collaboration honestly lack the tools to do so: no current framework can distinguish between a researcher who originated a research direction and one who adopted a direction proposed by AI, or between a researcher who critically evaluated AI-generated alternatives and one who accepted AI output without independent assessment. This gap is not a matter of compliance detail; it is a failure to capture the cognitive dynamics that determine what kind of intellectual contribution a paper actually represents.

We propose PAIRED — Process-Anchored Interaction Reporting for AI-Enabled Discovery — a dual-facing framework that addresses this gap through four design principles: process orientation, which takes the decision point rather than the research product as the fundamental unit of documentation; dual-facing output, which derives a structured publisher disclosure from a prospective author log without double work; decision-point granularity, which operates between session-level coarseness and message-level impracticality; and artifact-triggered logging, which provides an auditable rule against selective omission. We demonstrate PAIRED through worked examples, discuss its limitations openly, and propose a model-assisted adoption pathway that embeds the framework's logging discipline directly into AI research platforms.
\end{abstract}


\section{Introduction}\label{sec1}
The integration of generative AI into scientific research has outpaced the community's ability to describe it accurately. Across disciplines, authors now routinely collaborate with large language models at multiple stages of the research pipeline — from literature synthesis and hypothesis formation to methodological design, data analysis, and manuscript preparation. Yet the mechanisms available for disclosing this collaboration remain remarkably primitive. Most journal policies reduce AI involvement to a single declarative sentence appended to the acknowledgment section, affirming that AI was used and that the authors take responsibility for the output. More detailed guidelines, where they exist, instruct authors to report the "functional role" of AI at each stage of the research pipeline — a formulation that, while directionally correct, provides no operative definition of what a role is, how granularly it should be characterized, or what evidence should support the characterization.

The inadequacy of these mechanisms becomes visible at the moment a researcher attempts to document their collaboration honestly. Consider a concrete illustration: a researcher conducts an extended dialogue with a language model during the ideation phase of a study. The model misinterprets a key term, and the researcher's act of correcting that misinterpretation forces them to articulate — with a precision they had not previously achieved — the original direction of their own thinking. The idea that emerges is genuinely novel, entirely the researcher's own, and the model's contribution was, paradoxically, its misunderstanding. How should this be disclosed? No existing framework provides the vocabulary to distinguish this moment from one in which the model straightforwardly proposed an idea that the researcher adopted. Yet the epistemic difference between the two is precisely what the scientific community needs to understand in order to evaluate the integrity of the work.

Faced with this gap, a researcher might naturally turn to saving screenshots of their AI conversations — a practice that reveals, more clearly than any policy document, the nature of what is missing. Screenshots are an attempt to preserve process evidence in the absence of a process framework. But they are inconsistent across researchers, unscalable across papers, and — critically — incomplete as documentation: a screenshot captures the exchange but not the judgment, the adoption but not the reasoning, the conversation but not the cognitive ownership. A reader examining a screenshot cannot determine whether the researcher originated an idea that the model elaborated, or adopted an idea that the model originated. The screenshot preserves the surface of the interaction while the epistemically significant layer — who thought what, and who decided — remains invisible.

Existing proposals for structured AI disclosure share a common architectural limitation: they are output-oriented. They document what AI produced — a literature summary, a code segment, a draft paragraph — rather than how the research actually unfolded. Frameworks such as PaperCard \cite{cho2023papercard} and taxonomies extending the CRediT \cite{holcombe2019contributorship} contributor model map AI involvement onto research products, not research process. This is understandable as a starting point, but it is insufficient for the purpose of scientific transparency. The products of a collaboration do not reveal its dynamics. Knowing that AI contributed to "methodology" does not tell a reader whether the researcher evaluated AI-generated alternatives and selected among them, or whether the researcher proposed a direction and used AI to implement it, or whether the model introduced a concept the researcher had not considered and the researcher adopted it wholesale. These are meaningfully different situations with meaningfully different implications for how the work should be read and cited.

This paper proposes PAIRED — Process-Anchored Interaction Reporting for AI-Enabled Discovery — a framework designed to address this gap. PAIRED shifts the unit of documentation from the research product to the decision point: a moment within the research process where a direction was chosen, an idea was adopted or rejected, or a methodological commitment was made. At each decision point, three dimensions are documented: origination (who seeded the idea), elaboration and evaluation (how it was developed and how alternatives were filtered), and direction (who made the final call and what artifact resulted). These entries are captured in a three-field micro-log triggered not by session boundaries but by artifact adoption — the moment something from an AI interaction enters the research in a form that will influence the final paper. The log serves dual purposes: as a prospective author record kept during research, and as the source from which a structured publisher disclosure is derived, eliminating the need for retrospective reconstruction.

PAIRED makes three contributions. First, it establishes the process-versus-output distinction as the foundational theoretical reorientation needed in AI disclosure practice. Second, it proposes a minimal viable framework — specific enough to be actionable, flexible enough to accommodate different research workflows and documentation media. Third, it demonstrates the framework through three worked examples drawn from a real, published preprint, showing how PAIRED applies retrospectively to moments of human-AI collaboration that existing disclosure frameworks cannot adequately characterize. Notably, the development of PAIRED is itself documented using the framework's own micro-log — a reflexive application that the authors regard not as a rhetorical flourish but as a necessary demonstration of the framework's validity under the conditions of its own creation.

\section{Background and Related Work}\label{sec2:litrev}

The disclosure of AI involvement in scientific research has attracted growing attention from publishers, professional bodies, and the research community since the public release of capable generative language models in late 2022. The resulting landscape of policies and proposals can be organized into three layers: institutional publisher requirements, structural disclosure frameworks, and taxonomic extensions to existing contributor models. Across all three layers, a shared architectural assumption persists — one that PAIRED is designed to challenge.

\subsection{Institutional Publisher Requirements}

Major publishers and editorial bodies have moved toward mandatory AI disclosure at varying levels of specificity. The International Committee of Medical Journal Editors (ICMJE) \cite{kumar2026icmje} now requires explicit disclosure of AI use in drafting, editing, translation, image generation, and data analysis, while affirming that AI cannot be listed as an author and that human authors bear full responsibility for AI-assisted content. JAMA has developed one of the more detailed disclosure frameworks among major journals \cite{alfayyad2026self}, requiring authors to specify the AI tools used, the tasks for which they were used, and a confirmation that outputs were verified by the authors. The Committee on Publication Ethics (COPE) has actively debated the boundaries of required disclosure \cite{COPE2024AuthorshipAI}, including questions about what levels of AI involvement are ethically significant, how disclosure standards can be enforced at submission, and how policies can remain responsive to a technology that continues to evolve rapidly.

Despite this activity, the resulting requirements share a common structural form: they are checklist-based and product-oriented. An author affirms that AI was used, names the tool, identifies the research phase, and confirms that the output was reviewed. What is not captured — and what no current institutional requirement asks for — is the process by which AI involvement shaped the research: which directions were considered and rejected, whose judgment filtered the alternatives, and at what moments the human researcher's thinking was substantively altered by the interaction. The disclosure that results is accurate in a narrow sense but uninformative in the sense that matters most for scientific transparency.

\subsection{Structural Disclosure Frameworks}

Several research proposals have attempted to move beyond checklist compliance toward structured disclosure. PaperCard \cite{cho2023papercard}, introduced in 2023, proposes a standardized declaration comprising four components that authors insert into a designated section of the manuscript, covering the nature of AI use, the tools involved, and a statement of author responsibility. Its companion tool, Cardwriter, automates the generation of these declarations from author input, producing structured text suitable for inclusion in acknowledgment sections or supplementary materials. While PaperCard represents a meaningful advance over unstructured prose disclosure, its components remain organized around the output of AI use — what the tool produced — rather than the process through which that output was generated and adopted.

The Generative AI Delegation Taxonomy (GAIDeT), published in 2025 \cite{suchikova2026gaidet}, offers a more granular approach by providing a vocabulary for the types of tasks that researchers delegate to generative AI across the scientific research and publishing workflow. GAIDeT is a notable contribution because it acknowledges that delegation is a meaningful unit of analysis — that the act of assigning a task to AI carries disclosure significance. However, delegation is still an output-adjacent concept: it describes what was handed over, not what happened at the boundary where human judgment determined whether to adopt, modify, or reject what came back.

A complementary line of work has proposed decomposing the research process into stages as a scaffold for AI disclosure. A recent survey of AI scientists \cite{tie2025survey} proposes a six-stage methodological framework — literature review, idea generation, experimental preparation, experimental execution, scientific writing, and paper generation — as a structure within which AI involvement can be located and described. This stage-based approach is directionally aligned with PAIRED's design, and the present framework draws on it in constructing its four-stage research pipeline. However, stage-level disclosure remains coarse: knowing that AI was involved in "idea generation" does not reveal whether the researcher evaluated AI-generated alternatives and selected among them, proposed a direction and used AI to pressure-test it, or encountered an AI misinterpretation that forced a clarification that itself became the contribution.

\subsection{The Shared Limitation and the Gap PAIRED Addresses}

Taken together, the existing landscape — institutional requirements, structural frameworks, and stage-based taxonomies — shares a foundational architectural assumption: that the appropriate unit of AI disclosure is the research product. A draft paragraph, a literature summary, a methodological choice, a figure — these are the entities that existing frameworks ask authors to tag with AI involvement. This product orientation is understandable as a regulatory starting point, because products are visible, bounded, and verifiable in a way that process is not.

But it produces a systematic blind spot. The epistemically significant questions in human-AI research collaboration are not about products; they are about the cognitive dynamics that produced them. Who originated the idea that became the paper's central contribution? When AI proposed an alternative the researcher had not considered, what judgment did the researcher apply in evaluating it? When the researcher used AI to pressure-test a position they already held, and the AI's challenge sharpened rather than changed their thinking, how should that interaction be characterized? These questions cannot be answered by examining outputs. They require a framework oriented toward process — specifically, toward the decision points within that process where cognitive ownership, intellectual origination, and evaluative judgment can be documented with precision.

It is this gap that PAIRED addresses. The following section details the framework's architecture, its theoretical grounding in the process-versus-output distinction, and the practical mechanisms through which it enables both prospective author documentation and structured publisher disclosure.

\section{Methods}\label{sec3:methods}

PAIRED — Process-Anchored Interaction Reporting for AI-Enabled Discovery — is a dual-facing framework for documenting the role of generative AI in scientific research. It is designed to serve two audiences simultaneously: researchers, who need a practical discipline for recording AI involvement as it happens, and publishers, who need a structured, comparable disclosure at the point of submission. The framework's central design commitment is that both audiences are served by the same underlying record — the author log — from which the publisher disclosure is derived. This eliminates the burden of retrospective reconstruction and ensures that what is disclosed reflects what was actually documented during the research process.

\begin{figure}[ht]
  \centering
  \includegraphics[width=\linewidth]{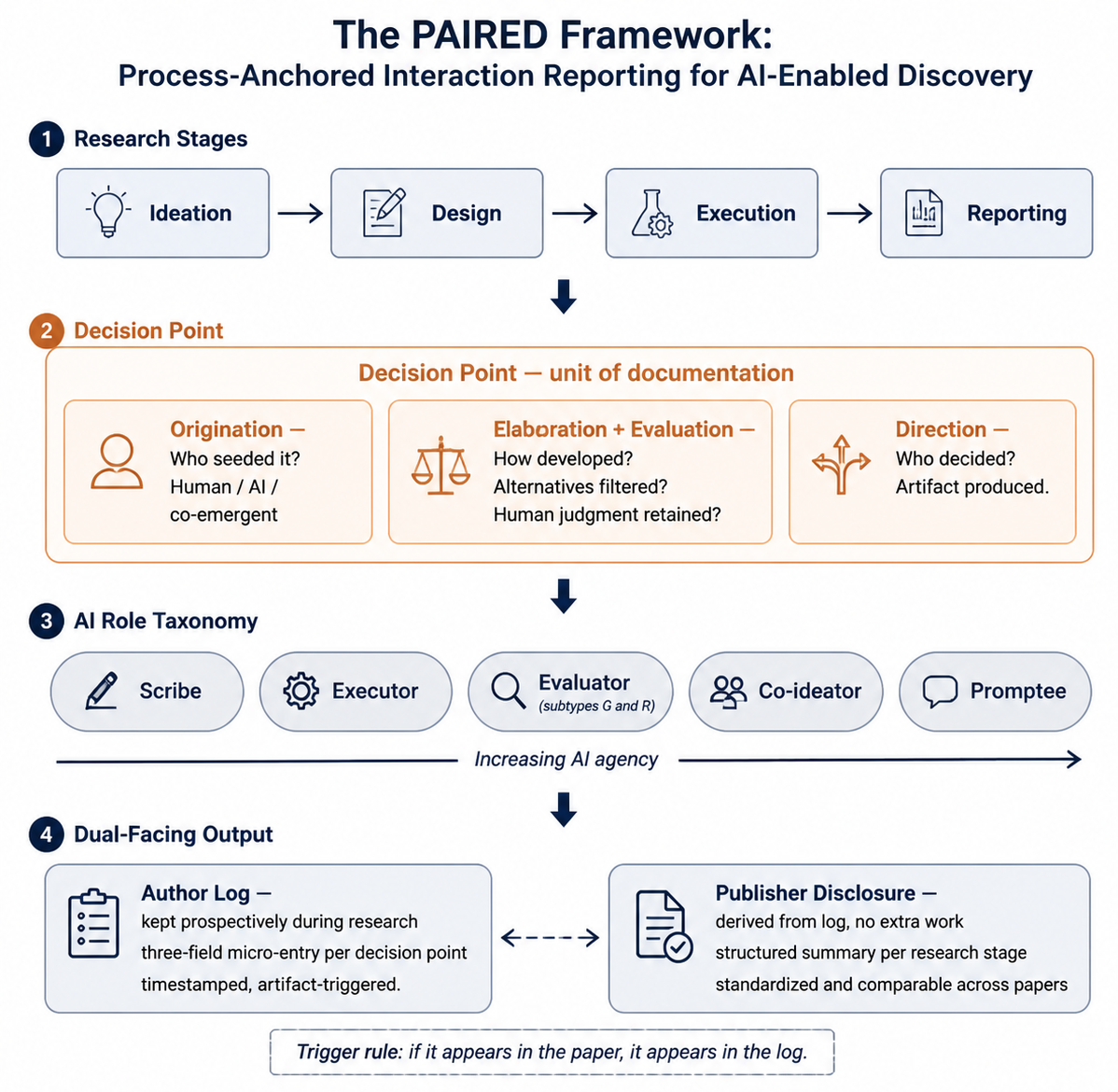}
  \caption{\textbf{The PAIRED Framework}. Four-layer architecture showing research stages, the decision point as the unit of documentation, the AI role taxonomy arranged by agency level, and the dual-facing output structure.}
  \label{fig:the_paired_framework}
\end{figure}

PAIRED is organized around four components, each addressed in a dedicated subsection below: the design principles that govern its architecture, the research stage structure within which documentation is located, the decision point as the fundamental unit of documentation, and the AI role taxonomy that provides the controlled vocabulary for characterizing AI involvement. The four-layer architecture of the proposed framework is depicted in Fig.~\ref{fig:the_paired_framework}.

\subsection{Design Principles}

The four principles that govern the architecture of PAIRED and distinguish it from existing disclosure approaches are the following.

\textit{\textbf{Process orientation}}. The primary unit of documentation in PAIRED is not a research product but a decision point — a moment within the research process where a direction was chosen, an idea was adopted or rejected, or a methodological commitment was made. This orientation is the framework's foundational departure from existing approaches, all of which document what AI produced rather than how the research unfolded. Process orientation does not preclude product documentation; it subsumes it. Every decision point produces an artifact, and that artifact is one of the three fields documented in the micro-log entry. But the artifact is recorded as the outcome of a decision, not as the primary unit of analysis.

\textit{\textbf{Dual-facing output}}. PAIRED produces two views of the same underlying record. The author log is a prospective, private record kept during research — structured, timestamped, and artifact-triggered. The publisher disclosure is a structured summary derived from the log at the point of manuscript submission, organized by research stage. Because the disclosure is derived rather than separately composed, it does not require the author to reconstruct their process after the fact, and it cannot diverge from the log without that divergence being detectable. Publishers who adopt PAIRED as a submission standard may request the log itself, the derived disclosure, or both.
    
\textit{\textbf{Decision-point granularity}}. PAIRED operates at a level of granularity between session-level logging, which is too coarse and difficult to implement consistently, and message-level logging, which is too fine and impractical for most research workflows. The decision point is the natural unit because it corresponds to a moment of human agency: the moment a researcher acts on something that emerged from an AI interaction. This granularity is sufficient to capture the epistemically significant dynamics of human-AI collaboration — origination, evaluation, and direction — without imposing an unsustainable documentation burden.

\textit{\textbf{Artifact-triggered logging}}. The trigger for a log entry in PAIRED is not a session boundary, a time interval, or a subjective assessment of significance. It is the adoption of an artifact: the moment something from an AI interaction enters the research in a form that will influence the final paper. This trigger is both objective and auditable. It produces a natural rule against selective omission: if an artifact appears in the paper, a log entry for the decision point that produced it must exist. A reviewer examining the final paper can, in principle, trace every named element — every research question, methodological choice, structural decision, or coined term — to a corresponding log entry.

\subsection{Research Stages}
PAIRED organizes the research process into four stages: Ideation, Design, Execution, and Reporting. These stages provide the structural scaffold within which decision points are located, and they serve as the organizing categories for the publisher disclosure.

\textit{Ideation} encompasses the early activities of a research project: literature engagement, gap identification, hypothesis formation, and the crystallization of the research question. AI involvement at this stage most commonly takes the form of literature synthesis, brainstorming, and the pressure-testing of nascent ideas. It is also the stage at which the epistemically most significant interactions tend to occur — where the boundary between the researcher's original thinking and AI-influenced thinking is most consequential and most difficult to characterize without a process-oriented framework.

\textit{Design} encompasses the translation of a research question into a methodological plan: the selection of methods, frameworks, instruments, and analytical approaches. AI involvement at this stage may include the generation and evaluation of architectural alternatives, the proposal of specific tools or structures, and the elaboration of implementation plans. The design stage is where AI contributions are most likely to be adopted in a form that is directly visible in the final paper — as named components, structural choices, or methodological commitments.

\textit{Execution} encompasses the implementation of the research plan: data collection, analysis, synthesis, and the interpretation of results. AI involvement at this stage may include code generation, analytical assistance, the organization of outputs, and the synthesis of findings across sources. Execution-stage AI involvement is often the most straightforwardly tool-like — AI as executor of a human-specified plan — but it can also include moments of genuine co-ideation where the interpretation of results is shaped by AI-generated framings.

\textit{Reporting} encompasses the translation of research findings into a manuscript: structuring, drafting, revising, and refining. AI involvement at this stage is the most widely disclosed under current publisher requirements, and the most easily characterized under existing frameworks, because it maps most directly onto identifiable textual products. PAIRED does not exclude reporting-stage disclosure but treats it as one stage among four rather than as the primary site of disclosure concern.

The four-stage structure is deliberately coarser than some existing proposals, which decompose the research process into five or six stages. This choice reflects a design priority: the stages serve as organizational containers for decision points, not as units of analysis in themselves. Finer stage granularity does not improve disclosure quality if the decision point — the actual unit — is documented with precision.

\subsection{The Decision Point}

The decision point is the fundamental unit of documentation in PAIRED. A decision point is defined as any moment in the research process at which a direction was chosen, an idea was adopted or rejected, or a methodological commitment was made, and at which AI involvement was present in a form that influenced the outcome. Not every AI interaction constitutes a decision point; routine interactions that produce nothing adopted into the research do not require documentation. The artifact-trigger rule operationalizes this distinction: a decision point requiring documentation is one that produced an artifact that appears in the final paper. Each decision point is documented along the following three dimensions.

\textit{Origination} captures who seeded the idea, concept, or direction that the decision point resolved. The origination dimension distinguishes three conditions: human-originated, where the researcher introduced the core idea and AI elaborated or implemented it; AI-originated, where the model introduced a concept or direction the researcher had not formulated and the researcher adopted it; and co-emergent, where the idea arose from the interaction itself in a way that cannot be cleanly attributed to either party. Origination is the dimension most directly relevant to questions of intellectual authorship and is the dimension that existing frameworks are least equipped to capture.

\textit{Elaboration and evaluation} captures how the idea was developed and how alternatives were filtered. This dimension documents whether AI generated alternatives that the researcher evaluated, whether the researcher generated alternatives that AI assessed, and — critically — whether human judgment was retained as the filtering mechanism or whether the researcher deferred to AI assessment without independent evaluation. The elaboration and evaluation dimension is treated as a single dimension in PAIRED because in practice these two activities are inseparable: the development of an idea and the judgment of its alternatives occur in the same dialogic space and cannot be cleanly sequenced.

\textit{Direction} captures who made the final call and what artifact resulted. The direction dimension documents whether the decisive judgment was the researcher's own, whether it was substantially shaped by AI recommendation, or whether the researcher deferred to AI output without independent assessment. It also records the artifact produced — the named element, structural choice, or methodological commitment that enters the paper as a result of the decision point. The artifact field in the direction dimension is the anchoring element that connects the log entry to the final paper and enables the artifact-trigger auditing rule.

These three dimensions are captured in a three-field micro-log entry, which constitutes the practical implementation of PAIRED's author-side log. The three fields are:

\begin{itemize}
    \item Artifact adopted: A one-sentence description of what entered the research as a result of this decision point.

    \item Role label: The AI role taxonomy label (see Section 3.4) that best characterizes AI's contribution at this decision point.

    \item Human judgment applied: A one-sentence description of what the researcher accepted, rejected, or modified, and on what basis.
\end{itemize}
The micro-log entry is intentionally minimal. Three fields, each capped at one sentence, impose a discipline of precision without creating a documentation burden that would discourage prospective use. The entries are kept in whatever medium the researcher already uses for research notes — a running document, a reference manager, a dedicated log file — because \textit{PAIRED is meant to specify the fields and the trigger, not the medium}.

\subsection{The AI Role Taxonomy}
PAIRED provides a five-role taxonomy for characterizing AI's contribution at each decision point. The roles are arranged along a single dimension of increasing AI agency — from roles in which AI operates entirely within human-specified parameters to roles in which AI substantively shapes the direction of the research. The taxonomy is applied at the level of the individual decision point, not at the level of the research stage or the paper as a whole, reflecting the empirical reality that AI's role typically varies across the research process and cannot be accurately summarized by a single label. The five roles, from lowest to highest AI agency, are as follows.

\textit{\textbf{Scribe}}. AI operates as a language processor, polishing, formatting, paraphrasing, or translating content that is entirely human-specified in substance. No new ideas, structures, or directions enter the research through AI's contribution. The human's intellectual content is preserved; AI improves its surface form. This is the role most commonly disclosed under current publisher requirements and the role that existing frameworks handle most adequately.

\textit{\textbf{Executor}}. AI implements a human-specified plan, generating code, running analyses, producing outputs, or completing tasks whose parameters were entirely determined by the researcher. The researcher's judgment governs what is done; AI's contribution is the doing of it. Executor-role contributions may be substantial in terms of labor but are low in terms of intellectual agency — the research direction is not influenced by what AI produces, only by what the researcher specified.

\textit{\textbf{Evaluator}}. AI assesses alternatives, pressure-tests positions, or provides critical feedback on researcher-generated ideas or plans. The Evaluator role has two subtypes that PAIRED treats as distinct. Evaluator-G (generative) denotes interactions in which AI generates alternatives that it then assesses, with the researcher choosing among AI-produced options. Evaluator-R (reflexive) denotes interactions in which the researcher already holds a position and uses AI to stress-test it — where AI's contribution is challenge and sharpening rather than generation. The distinction between these subtypes is epistemically significant: in Evaluator-G interactions, AI contributes to the option space; in Evaluator-R interactions, AI contributes to the refinement of the researcher's own thinking without expanding the option space.

\textit{\textbf{Co-ideator}}. AI introduces a concept, structure, term, or direction that the researcher had not formulated, and the researcher adopts it — with or without modification — as part of the research. Co-ideator interactions are the most epistemically complex because they involve a genuine transfer of intellectual content from AI to the research, mediated by the researcher's judgment of adoption. The researcher's evaluation and decision to adopt remain the operative human contribution, but the origination of the adopted element lies with AI. Co-ideator is the role that existing frameworks most systematically mischaracterize, either by overstating AI's contribution (treating adoption as co-authorship) or understating it (treating it as mere tool use).

\textit{\textbf{Promptee}}. AI substantially shapes the direction of the research through outputs that the researcher follows without independent critical evaluation. The Promptee role represents the highest AI agency in the taxonomy and the lowest exercise of researcher judgment at the decision point. It is included in the taxonomy not as a normative category — PAIRED does not prescribe how much AI agency is appropriate — but as a disclosure category, enabling researchers and readers to identify moments where human oversight was minimal. Transparent disclosure of Promptee-role interactions is among the most important functions PAIRED serves.

The taxonomy is applied by the researcher at the time of logging, based on their own characterization of the interaction. This reliance on researcher self-report is a limitation acknowledged in Section 5.3. It is partially mitigated by the artifact-trigger rule, which ensures that every adopted artifact has a corresponding log entry, and by the specificity of the role definitions, which reduce the ambiguity that makes self-report unreliable in less structured frameworks.

\section{Worked Examples}\label{sec:worked_examples}

The following examples apply PAIRED retrospectively to three decision points drawn from the development of a published preprint on AI-assisted swimming coaching --- a multimodal dataset synthesis framework for RAG-based athletic coaching systems \cite{alkabbany2026synthesizing}. The preprint was selected because its development involved substantive human-AI collaboration across multiple research stages, and because one of its authors maintained sufficiently detailed recollections of key interactions to permit accurate retrospective documentation. The examples are presented in the order in which the decision points occurred during the research process.

Retrospective application of PAIRED is less reliable than prospective logging, for reasons discussed in Section~\ref{sec:limitations}. These examples are offered as demonstrations of the framework's analytical vocabulary and its capacity to distinguish epistemically different forms of human-AI collaboration --- not as a gold standard of the documentation practice PAIRED recommends. Prospective application, triggered at the moment of artifact adoption, remains the framework's intended mode of use.

It is worth mentioning that the third example documents a decision point from the development of PAIRED itself. The author regards this not as a rhetorical device but as a necessary demonstration --- a framework for documenting human-AI collaboration that cannot account for its own creation would be incomplete. The micro-log entry for that decision point was kept prospectively during the framework development process and is reproduced here without modification.

In each of the following decision points, we present both a micro-log table and a prose rendering. This dual presentation is intentional: it demonstrates that PAIRED is flexible with respect to the exact form of documentation, accommodating structured tabular records and narrative prose equally, provided that the decision point remains the unit of reporting and the three fields — artifact adopted, role label, and human judgment applied — are preserved in substance.

\subsection{Decision Point 1 — The RAG Synthesis Redirection (Ideation Stage)}

\textbf{Context.} During an extended literature review conversation with a language model, the researcher was exploring gaps in AI applications for swimming coaching. The model repeatedly used the phrase ``RAG synthesis'' to mean the architectural design of a retrieval system --- the engineering of retrieval logic and code. The researcher, for whom ``synthesis'' carried an unambiguous generative meaning --- the \textit{creation} of a dataset using LLMs --- recognized the misalignment and explicitly corrected it. The act of correcting the model's interpretation forced a precision of articulation that crystallized the research direction: not building a RAG system, but synthesizing the expert knowledge corpus that would ground one.

\textbf{Micro-log entry.} A demonstration of the micro-log entry that corresponds to decision point 1 is shown in Table~\ref{tab:dp1}.

\begin{table}[h]
\caption{PAIRED micro-log entry for Decision Point 1 (table format).}
\centering
\renewcommand{\arraystretch}{1.5}
\begin{tabular}{>{\bfseries}p{2.8cm} p{9.5cm}}
\hline
Field & Entry \\
\hline
Artifact adopted &
    Core research direction: RAG synthesis defined as the generative
    creation of an expert-level dataset using LLMs, distinct from
    RAG architectural design. \\
Role label &
    Evaluator-R --- AI misinterpreted a key term; the researcher's
    correction of that misinterpretation crystallized their own
    thinking. \\
Human judgment applied &
    Rejected AI's architectural interpretation of ``synthesis'';
    retained and sharpened own generative definition; redirected the
    entire research trajectory on the basis of that distinction. \\
\hline
\end{tabular}
\label{tab:dp1}
\end{table}

\textbf{Prose rendering.} During the ideation phase, an extended conversation with a language model around the concept of RAG synthesis revealed a terminological misalignment: the model consistently interpreted ``synthesis'' in the architectural sense, referring to the design of retrieval logic, whereas the researcher's intended meaning was generative --- the creation of a structured expert dataset using LLMs. The researcher's act of correcting this misinterpretation served as the clarifying event that crystallized the research direction. AI's role at this decision point is classified as Evaluator-R: the model contributed not by proposing the direction but by misunderstanding it, and the researcher's judgment in articulating the correction constitutes the intellectual act of origination.

\textbf{What PAIRED captures that existing frameworks miss.} This decision point would be invisible under any existing disclosure framework. No AI-generated artifact was adopted; no AI-proposed alternative was selected. The model's contribution was a misunderstanding, and the researcher's contribution was the correction of it. Yet this interaction was among the most consequential in the entire research process --- it produced the paper's central research direction. The Evaluator-R subtype is the only available label that accurately characterizes this dynamic, and it is a label that no existing taxonomy provides. A conventional disclosure would either omit this interaction entirely or misclassify it as routine AI assistance, in both cases obscuring the fact that the research direction was entirely human-originated.

\subsection{Decision Point 2 — The Golden Triplets Structure (Design Stage)}

\textbf{Context.} Having established the generative research direction, the researcher turned to the structural design of the synthesized corpus. The researcher's initial formulation was to organize the corpus as JSON objects with various metadata fields specifying domain, context, and other attributes. During a design conversation with a language model, the model proposed structuring the corpus as Question-Context-Answer triplets --- a format that imposed a retrieval-native structure on the knowledge base --- and coined the term ``Golden Triplets'' for the validated records. The researcher evaluated the proposal, recognized its superiority over generic JSON objects for RAG grounding purposes, and adopted both the structure and the term. Both appear in the final paper as named contributions of the methodology.

\textbf{Micro-log entry.} A demonstration of the micro-log entry that corresponds to decision point 2 is shown in Table~\ref{tab:dp2}.

\begin{table}[h]
\caption{PAIRED micro-log entry for Decision Point 2 (table format).}
\centering
\renewcommand{\arraystretch}{1.5}
\begin{tabular}{>{\bfseries}p{2.8cm} p{9.5cm}}
\hline
Field & Entry \\
\hline
Artifact adopted &
    Question-Context-Answer triplet structure as the organizational
    format for the synthesized corpus; the term ``Golden Triplets''
    for validated records. \\
Role label &
    Co-ideator --- AI proposed a specific architecture and coined a
    term the researcher had not formulated. \\
Human judgment applied &
    Accepted triplet structure as superior to generic JSON objects
    for RAG grounding; retained AI-coined term; integrated both into
    the methodology and the paper's identity. \\
\hline
\end{tabular}
\label{tab:dp2}
\end{table}

\textbf{Prose rendering.} During the design phase, the researcher's initial plan was to structure the synthesized corpus as JSON objects with multiple metadata fields. In conversation with a language model, the model proposed reframing the corpus structure as Question-Context-Answer triplets --- a format native to retrieval-augmented generation evaluation --- and introduced the term ``Golden Triplets'' to designate validated records. The researcher evaluated this proposal against the original JSON formulation and adopted both the structure and the term, recognizing the triplet format's advantages for RAG grounding and benchmarking. AI's role at this decision point is classified as Co-ideator: the model introduced a concept and a term the researcher had not formulated, and the researcher's judgment of adoption --- informed by domain understanding of RAG system requirements --- constitutes the operative human contribution.

\textbf{What PAIRED captures that existing frameworks miss.} This decision point is partially visible under existing frameworks --- the acknowledgment section of the published preprint notes that the triplet structure and the term ``Golden Triplets'' originated with the language model. But existing frameworks provide no vocabulary for characterizing what kind of contribution this represents, how it differs from the contribution in Decision Point 1, or what human judgment mediated the adoption. PAIRED's Co-ideator label, combined with the human judgment field, makes these distinctions explicit and comparable across papers. A reader can see not only that AI proposed the triplet structure but that the researcher evaluated it against a specific alternative and adopted it on the basis of domain-specific reasoning --- a level of transparency that prose acknowledgment alone cannot consistently achieve.

\subsection{Decision Point 3 — The PAIRED Framework Name (Design Stage)}

\textbf{Context.} During the development of the present framework, the researcher and AI collaborator discussed naming options for what was then an unnamed process-anchored disclosure framework. The researcher had initially proposed embedding the AI system's name in the framework title as a form of credit --- a gesture toward the collaborative nature of the framework's development. Following an honest discussion of the principled and practical objections to that approach, the researcher identified two non-negotiable semantic anchors for the name: the concept of \textit{process}, as the framework's genuine differentiator from existing approaches, and the concept of \textit{generative AI}, as the specific collaboration partner the framework is designed to document. The AI collaborator generated a set of candidate acronyms satisfying both constraints. The researcher selected PAIRED --- Process-Anchored Interaction Reporting for AI-Enabled Discovery --- from the candidate set on the basis of semantic fit, memorability, and the word's own meaning as a descriptor of human-AI collaboration.

\textbf{Micro-log entry.} A demonstration of the micro-log entry that corresponds to decision point 3 is shown in Table~\ref{tab:dp3}.

\begin{table}[h]
\caption{PAIRED micro-log entry for Decision Point 3 (table format).}
\centering
\renewcommand{\arraystretch}{1.5}
\begin{tabular}{>{\bfseries}p{2.8cm} p{9.5cm}}
\hline
Field & Entry \\
\hline
Artifact adopted &
    Framework name: PAIRED --- Process-Anchored Interaction Reporting
    for AI-Enabled Discovery. \\
Role label &
    Co-ideator --- researcher specified naming constraints; AI
    generated the candidate set; researcher selected from it. \\
Human judgment applied &
    Rejected initial naming approach on principled grounds;
    identified two non-negotiable semantic constraints (process,
    generative AI); selected PAIRED from AI-generated candidates as
    the strongest fit on grounds of semantic accuracy, memorability,
    and the word's own descriptive value. \\
\hline
\end{tabular}
\label{tab:dp3}
\end{table}

\textbf{Prose rendering.} During the design phase of the present framework, the naming of PAIRED itself constituted a documented decision point. The researcher identified two semantic requirements that any candidate name must satisfy: it must foreground \textit{process} as the framework's primary differentiator, and it must reference \textit{generative AI} as the specific form of collaboration the framework addresses. A candidate set was generated by the AI collaborator satisfying both constraints. The researcher evaluated the candidates and selected PAIRED --- Process-Anchored Interaction Reporting for AI-Enabled Discovery --- on the basis of semantic precision, memorability, and the appropriateness of the word ``paired'' as a descriptor of the human-AI collaborative relationship the framework is designed to document. AI's role at this decision point is classified as Co-ideator: the naming constraints were entirely human-specified, the candidate set was AI-generated, and the selection was entirely human-made.

\textbf{What PAIRED captures that existing frameworks miss.} This example is included not because the naming of a framework is an unusually significant research decision, but because it demonstrates PAIRED's reflexive validity --- the framework can document its own creation with the same precision it brings to any other human-AI collaborative research process. It also illustrates a consistent pattern across all three decision points: the researcher's intellectual contribution is most accurately characterized not by what AI produced, but by the judgment applied at the moment of adoption. In Decision Point~1, that judgment was a correction. In Decision Point~2, it was a domain-informed evaluation. In Decision Point~3, it was a constraint-specification followed by selection. These are three meaningfully different cognitive acts, and PAIRED is the only available framework that provides the vocabulary to distinguish them.

\section{Discussion}\label{sec:discussion}

The worked examples in Section~\ref{sec:worked_examples} demonstrate PAIRED's analytical vocabulary across three decision points that are, by any existing disclosure standard, indistinguishable. All three would be reported under current publisher requirements as something like: ``AI was used during the ideation and design phases of this research.'' PAIRED reveals that this single sentence conceals three fundamentally different epistemic situations --- a correction, a domain-informed adoption, and a constraint-specified selection --- each with different implications for how the research should be read and how the researcher's intellectual contribution should be understood. This section draws out the broader implications of that demonstration, addresses the framework's open questions, and acknowledges its limitations honestly.

\subsection{What PAIRED Captures That Checklists Cannot}

The central insight that motivates PAIRED is that the epistemically significant questions in human-AI research collaboration are not about products but about the cognitive dynamics that produced them. Checklists can capture what AI produced. They cannot capture who originated the idea that AI elaborated, what judgment the researcher applied when evaluating AI-generated alternatives, or whether the researcher's thinking was sharpened, redirected, or simply implemented by AI involvement. These distinctions matter for at least three reasons.

First, they bear on questions of intellectual authorship. The research community has reached a working consensus that AI cannot be an author --- it cannot take responsibility for the work, it cannot be held accountable for errors, and it does not have the standing that authorship confers. But this consensus, while correct, does not resolve the harder question: when AI originates an idea that a human researcher adopts, evaluates, and integrates into a paper, what exactly is the human's intellectual contribution? PAIRED provides the vocabulary to answer this question precisely rather than gesturally. The human's contribution is the judgment of adoption --- the domain-specific evaluation that determines whether the AI-generated idea is correct, useful, and worth integrating. That judgment is a genuine intellectual act, and it deserves to be documented as such rather than either inflated into co-authorship or deflated into mere tool use.

Second, they bear on the reproducibility and interpretability of research. A paper that reports ``AI assisted with methodology'' provides a reader with no basis for evaluating whether the methodological choices are the researcher's own considered judgments or AI-generated defaults that the researcher accepted without independent evaluation. PAIRED's direction dimension, and specifically the distinction between researcher-directed and AI-deferred decision points, gives readers the information they need to calibrate their confidence in the work's methodological integrity. A corpus of papers documented using PAIRED would, over time, enable the research community to develop empirically grounded norms about what levels and types of AI involvement are consistent with the standards of different research domains.

Third, they bear on the integrity of the historical record. Science depends on the accurate attribution of ideas to their originators --- not for reasons of vanity but because attribution enables the tracing of intellectual lineage, the identification of priority, and the correction of errors at their source. A disclosure framework that cannot distinguish between a researcher who originated a research direction and a researcher who adopted one proposed by AI is a framework that corrupts the historical record it is designed to protect. PAIRED is designed to preserve that record with the precision the moment demands.

\subsection{Open Questions}

PAIRED is proposed as a minimal viable framework --- specific enough to be actionable, flexible enough to accommodate different research workflows. Several questions remain open and are identified here as directions for community refinement.

\textit{\textbf{Multi-author teams.}} The present framework is designed primarily for the single-researcher or lead-researcher use case. When multiple human authors collaborate on a paper, each interacting with AI tools independently, the question of how individual logs are reconciled into a shared disclosure is non-trivial. One author's Evaluator-R interaction and another's Co-ideator interaction during the same research phase may produce artifacts that are difficult to disentangle in the final paper. Future work should address the coordination protocols needed for multi-author PAIRED documentation, including whether a designated corresponding author should maintain the master log or whether individual logs should be submitted in parallel.

\textit{\textbf{AI tools without conversation logs.}} PAIRED's artifact-trigger rule assumes that AI involvement produces a traceable interaction --- a conversation, a session, a documented exchange. Many AI-assisted research tools do not produce such records: AI-assisted statistical software, automated literature screening tools, and AI-enhanced writing environments may influence the research without generating a retrievable log. The framework's current specification handles these cases through the researcher's own micro-log entry, which does not require a retrievable AI-side record. However, the auditability that the artifact-trigger rule provides in conversation-based interactions is weaker for tool-based interactions, and future work should address how PAIRED can be extended to cover the full range of AI tool types researchers now use.

\textit{\textbf{Publisher adoption and log submission.}} PAIRED proposes that the publisher disclosure is derived from the author log, but it does not specify whether publishers should request the log itself alongside the derived disclosure. There are arguments on both sides. Requesting the log increases transparency and enables reviewers to verify that the disclosure is accurately derived. It also raises questions about the privacy of research-in-progress notes and the practical burden of preparing a log for submission. A tiered approach --- in which the derived disclosure is required at submission and the underlying log is available on request during peer review --- may represent a workable compromise, but this is a question for the publishing community to resolve in dialogue with researchers.

\textit{\textbf{Standardization across disciplines.}} The four research stages and five role labels in PAIRED are designed to be domain-agnostic, but their application will inevitably vary across disciplines with different research workflows. A computational study, a qualitative ethnography, and a clinical trial involve sufficiently different processes that the decision points occurring within them may not map cleanly onto the same stage and role vocabulary. Community-led refinement of PAIRED for specific disciplinary contexts --- analogous to the domain-specific extensions that have developed around the CRediT taxonomy --- would strengthen the framework's practical utility and its capacity to generate comparable disclosures across the literature.

\subsection{Limitations}
\label{sec:limitations}

PAIRED has three limitations that the author regards as important to acknowledge explicitly, both in the interest of intellectual honesty and because they identify the directions in which the framework most needs development.

\textit{\textbf{Reliance on researcher self-report.}} The micro-log is completed by the researcher on the basis of their own characterization of the interaction. This reliance on self-report is a structural limitation that PAIRED shares with every other disclosure framework --- including, ultimately, the checklists it is designed to replace. The artifact-trigger rule partially mitigates this limitation by making selective omission auditable in principle: an artifact that appears in the paper without a corresponding log entry is a detectable inconsistency. But within each log entry, the accuracy of the role label and the honesty of the human judgment field depend on the researcher's integrity. PAIRED cannot enforce honesty; it can only make dishonesty more visible by providing a structured record against which the final paper can be checked.

\textit{\textbf{Retrospective application.}} The worked examples in Section~\ref{sec:worked_examples} apply PAIRED retrospectively to interactions that occurred before the framework existed. Retrospective documentation is less reliable than prospective logging because memory of the precise dynamics of an interaction --- who originated what, what alternatives were considered, what judgment was applied --- degrades over time and is susceptible to post-hoc rationalization. The authors have been transparent about this limitation in introducing the worked examples, and they note that the third example, which documents a decision point from the development of PAIRED itself, was logged prospectively and does not carry this limitation. Future empirical work validating PAIRED should prioritize prospective application and should assess the reliability of retrospective documentation against prospective records where both are available.

\textit{\textbf{Framework maturity.}} PAIRED is proposed as a first-generation framework designed to improve on the current state without waiting for perfection. Several of its components --- the boundary conditions of the five role labels, the handling of co-emergent origination, the coordination protocols for multi-author teams --- require further development and empirical testing before the framework can be recommended as a universal standard. The author invites the research community to treat PAIRED as a working proposal, to apply it in their own work, and to report the cases it handles well and the cases it does not. A framework that accurately describes the present state of human-AI research collaboration will necessarily be a framework that evolves as that collaboration evolves.

\subsection{Toward Beta Adoption: Model-Assisted Micro-Logging}

The primary adoption barrier for PAIRED in its manually implemented form is the interruption of research flow. Prospective logging --- the framework's recommended mode --- requires a researcher to pause at the moment of artifact adoption and complete a three-field entry. While the entry is intentionally minimal, the discipline of recognizing a decision point, classifying a role, and articulating a judgment in the moment of research introduces a cognitive context-switch that some researchers may find difficult to sustain consistently, particularly during the fluid, exploratory phases of ideation and design where the most epistemically significant interactions tend to occur.

A practical pathway toward broader adoption lies in the integration of PAIRED's logging discipline directly into the AI platforms researchers already use. We propose that language model providers consider implementing an opt-in PAIRED logging mode --- activated by the researcher at the outset of a research conversation, analogous to the activation of existing extended research functionalities --- in which the model automatically generates candidate micro-log entries at each moment it identifies as a potential decision point. The identification heuristic would be straightforward: any interaction in which the model produces an artifact that the researcher explicitly accepts, builds upon, or integrates into their working document constitutes a candidate decision point, and the model generates a three-field entry --- artifact adopted, role label, human judgment applied --- based on its record of the interaction.

This mode does not transfer logging responsibility from the researcher to the model. It transfers the bookkeeping labor while preserving the researcher's epistemic authority. The model-generated log would be presented to the researcher as a draft record at the close of the conversation or at designated review intervals, and the researcher would be responsible for reading each entry thoroughly, modifying any entry that inaccurately characterizes the interaction, and approving the final log before it is submitted alongside the manuscript. Crucially, any modification the researcher makes to a model-generated entry would itself be flagged in the record --- creating a two-layer log that distinguishes what the model observed and what the researcher judged. This two-layer structure makes the auto-logging mode potentially more transparent than manual logging, not less, because it renders the researcher's evaluative acts visible as a distinct layer rather than embedding them invisibly in a self-reported record.

The proposed integration pattern has implications beyond individual researcher workflow. At the platform level, it represents a concrete specification for how AI providers can contribute to the emerging infrastructure of research transparency --- not by imposing disclosure requirements on researchers, but by making compliance with those requirements a natural byproduct of the research process itself. At the policy level, it suggests a pathway toward verified disclosure: a log generated and timestamped by the model during the research process carries an evidentiary weight that a retrospectively composed disclosure document cannot match. Publishers who adopt PAIRED as a submission standard could, in principle, request model-generated logs as a verification layer alongside the researcher-reviewed disclosure.

We offer this proposal as a call-to-action to AI platform developers, research tool providers, and the publishers whose submission standards shape researcher behavior. The technical implementation is within reach of any platform that already maintains conversation logs and supports extended research modes. The normative framework --- what to log, how to classify it, and who bears responsibility for its accuracy --- is what PAIRED provides. The combination of the two would represent a meaningful step toward a research ecosystem in which AI transparency is not an afterthought appended to a finished manuscript, but a discipline embedded in the research process from its first moment.

\section{Conclusion}

In this paper, we proposed PAIRED --- a Process-Anchored Interaction Reporting framework for AI-Enabled Discovery --- as a new approach to documenting AI contributions in scientific research. PAIRED is motivated by two complementary sources of insight: the systematic gaps in existing disclosure frameworks, which are uniformly output-oriented and therefore structurally unable to capture the dynamics of the research process, and the concrete reporting challenges encountered during the preparation of a recent preprint on AI-assisted swimming coaching, whose acknowledgment section revealed both the researcher's instinct for honest disclosure and the absence of a framework adequate to support it. Together, these two sources of motivation converge on a single diagnosis: what is missing from the current landscape of AI disclosure is not more detailed checklists but a fundamentally different unit of analysis.

PAIRED proposes the decision point as that unit, documenting each along three dimensions --- origination, elaboration and evaluation, and direction --- and classifying AI's contribution using a five-role taxonomy arranged along an agency gradient. This process-focused orientation is what fundamentally differentiates PAIRED from every existing reporting framework. Knowing what AI produced is not the same as knowing how the research unfolded, and knowing that AI assisted with methodology is not the same as knowing whether the researcher originated the methodological direction, evaluated AI-generated alternatives, or adopted an AI proposal without independent critical assessment. PAIRED is the first framework designed to capture these distinctions systematically, through a dual-facing architecture in which a prospective author log --- triggered by artifact adoption and kept in any medium --- serves as the source from which a structured publisher disclosure is derived.

The worked examples in Section~\ref{sec:worked_examples} demonstrate that PAIRED's analytical vocabulary successfully distinguishes epistemically different forms of human-AI collaboration that existing frameworks treat as identical, while its micro-log format remains sufficiently minimal for prospective use. Looking ahead, the model-assisted micro-logging proposal in Section~\ref{sec:discussion} outlines a concrete adoption pathway in which bookkeeping labor is absorbed by the AI platform while the researcher retains full epistemic authority over the final record. PAIRED is offered not as a finished standard but as a first-generation framework designed to improve on the current state without waiting for perfection, and the authors regard the community's engagement with its open questions as a continuation, rather than a conclusion, of the work this paper begins.

\section*{Acknowledgments}

The development of PAIRED involved substantive human-AI collaboration
across multiple stages of the research process. In the interest of
modeling the disclosure practice the framework proposes, the authors
document below all decision points from this work that meet PAIRED's
artifact-trigger criterion. Claude (Sonnet 4.6, Anthropic) was the
AI collaborator throughout the development of this framework.

\begin{table}[h]
\centering
\renewcommand{\arraystretch}{1.6}
\caption{PAIRED micro-log: decision points DP-I0 through DP-D3
from the development of this framework (part 1 of 4).}
\label{tab:acknowledgments1}
\begin{tabular}{>{\bfseries}p{1.1cm} >{\bfseries}p{1.8cm} p{3.8cm} p{3.8cm} p{3.8cm}}
\hline
\# & Stage & Artifact adopted & Role label & Human judgment applied \\
\hline

DP-I0 & Ideation &
    Recognition of a gap in existing AI disclosure frameworks and
    the decision to develop a structured alternative, originating
    from the author's own reporting experience and
    dissatisfaction with available tools. &
    None --- human-originated. &
    Voiced the concern, identified the need, and initiated the
    collaborative development process that produced PAIRED. \\

DP-I1 & Ideation &
    Process-oriented versus product-oriented as the foundational
    theoretical framing of PAIRED. &
    Co-ideator --- the diagnostic instinct was human; the precise
    theoretical framing emerged collaboratively. &
    Adopted the process-versus-output distinction as the framework's
    core differentiating claim after recognizing it accurately
    captured the insufficiency of existing approaches. \\

DP-I2 & Ideation &
    Literature landscape: CRediT, PaperCard, GAIDeT, and the
    six-stage survey framework as the relevant prior work. &
    Executor --- author specified the task; AI surfaced the
    relevant prior work; author determined relevance and
    framing. &
    Evaluated the returned literature against the framework's
    diagnostic argument; selected the works that most clearly
    illustrated the shared output-oriented limitation; structured
    Section~2 accordingly. \\

DP-D1 & Design &
    Dual-facing framework orientation: a single underlying log
    serving both an author-side record and a publisher-side
    disclosure. &
    Co-ideator --- the requirement for a framework serving both
    audiences was human-specified; the dual-facing architecture
    emerged collaboratively. &
    Adopted the dual-facing structure after recognizing it resolved
    the double-work problem that would otherwise undermine
    prospective adoption. \\

DP-D2 & Design &
    Decision-point granularity as the unit of documentation,
    positioned between session-level coarseness and message-level
    impracticality. &
    Co-ideator --- the dissatisfaction with existing granularity
    options was human-specified; the decision point as the natural
    unit emerged collaboratively. &
    Adopted decision-point granularity after evaluating it against
    session-level and message-level alternatives and recognizing it
    as the level that corresponds to a moment of human agency. \\

\hline
\end{tabular}
\end{table}

\begin{table}[h]
\centering
\renewcommand{\arraystretch}{1.6}
\caption{PAIRED micro-log: decision points DP-D3 through DP-D7
from the development of this framework (part 2 of 4).}
\label{tab:acknowledgments2}
\begin{tabular}{>{\bfseries}p{1.1cm} >{\bfseries}p{1.8cm} p{3.8cm} p{3.8cm} p{3.8cm}}
\hline
\# & Stage & Artifact adopted & Role label & Human judgment applied \\
\hline

DP-D3 & Design &
    Artifact-trigger rule: the moment of artifact adoption as the
    logging trigger rather than a session boundary or time interval. &
    Co-ideator --- the inadequacy of session boundaries as a trigger
    was human-identified; the artifact-adoption trigger emerged
    collaboratively. &
    Adopted the artifact-trigger rule after recognizing it resolved
    the session-definition problem and provided an objective,
    auditable logging criterion. \\

DP-D4 & Design &
    The ``if it's in the paper, it's in the log'' rule as the
    operative anti-omission mechanism. &
    Co-ideator --- the selective omission problem was
    human-identified; the artifact-based anti-omission rule emerged
    collaboratively. &
    Adopted the rule after recognizing it provided a structural
    rather than subjective criterion for completeness, making
    selective omission detectable in principle. \\

DP-D5 & Design &
    Five-role taxonomy arranged along an agency gradient: Scribe,
    Executor, Evaluator, Co-ideator, Promptee. &
    Co-ideator --- the need for a controlled vocabulary was
    human-specified; the five-role structure and agency gradient
    emerged collaboratively. &
    Adopted the five-role taxonomy after evaluating it against
    coarser and finer alternatives; retained the agency gradient as
    the organizing principle on the grounds that it captures the
    dimension most relevant to intellectual authorship questions. \\

DP-D6 & Design &
    Evaluator-R and Evaluator-G as subtypes of the Evaluator role,
    distinguishing reflexive stress-testing from generative
    alternative evaluation. &
    Co-ideator --- the inadequacy of a flat Evaluator label was
    human-identified from lived research experience; the R and G
    subtype distinction emerged collaboratively. &
    Adopted the subtype distinction after recognizing it captured
    an epistemically significant difference that a flat label would
    systematically obscure; chose subtype rather than separate role
    to preserve the five-role structure. \\

DP-D7 & Design &
    Three-field micro-log format: artifact adopted, role label,
    human judgment applied. &
    Co-ideator --- the need for a minimal structured entry was
    human-specified; the specific three-field structure emerged
    collaboratively. &
    Adopted the three-field format after evaluating it against
    fuller template alternatives; retained minimalism as a design
    priority on the grounds that documentation burden is the primary
    adoption barrier. \\

\hline
\end{tabular}
\end{table}

\begin{table}[h]
\centering
\renewcommand{\arraystretch}{1.6}
\caption{PAIRED micro-log: decision points DP-D8 through DP-E3
from the development of this framework (part 3 of 4).}
\label{tab:acknowledgments3}
\begin{tabular}{>{\bfseries}p{1.1cm} >{\bfseries}p{1.8cm} p{3.8cm} p{3.8cm} p{3.8cm}}
\hline
\# & Stage & Artifact adopted & Role label & Human judgment applied \\
\hline

DP-D8 & Design &
    Reflexive application: documenting PAIRED's own creation as a
    worked example, triggered by the author's observation that
    the framework development constituted a meta-analysis. &
    Co-ideator --- the meta-level observation was human-originated;
    the specific proposal to include PAIRED's own development as
    Decision Point 3 in Section~4 emerged collaboratively. &
    Adopted the reflexive application after recognizing it was not
    a rhetorical flourish but a necessary demonstration of the
    framework's validity under the conditions of its own creation. \\

DP-D9 & Design &
    Framework name: PAIRED --- Process-Anchored Interaction
    Reporting for AI-Enabled Discovery. &
    Co-ideator --- the two non-negotiable semantic constraints
    (process, generative AI) were human-specified; the candidate
    set was AI-generated; the selection was human-made. &
    Rejected an initial naming approach that embedded the AI
    system's name on principled grounds; identified semantic
    constraints; selected PAIRED from the candidate set on the
    basis of semantic precision, memorability, and the word's own
    descriptive value as a descriptor of human-AI collaboration. \\

DP-D10 & Design &
    Model-assisted micro-logging proposal: an opt-in platform-level
    integration in which the model generates candidate log entries
    and the researcher reviews, modifies, and approves them,
    producing a two-layer record. &
    Co-ideator --- the adoption barrier of interrupted research flow
    was human-identified; the specific platform integration pattern
    and the two-layer record structure emerged collaboratively. &
    Adopted the proposal after recognizing it resolved the primary
    adoption barrier without transferring epistemic authority from
    the researcher to the model; retained the two-layer structure
    as a transparency enhancement rather than a compliance
    mechanism. \\

DP-E1 & Execution &
    Screenshot from a conversation with Gemini (Google) as empirical evidence
    of the RAG synthesis redirection moment, serving as the primary
    source material for Decision Point 1 in Section~4. &
    None --- human-originated; the author supplied the
    empirical material independently. &
    Selected the screenshot as the most direct evidence of the
    origination dynamic that motivated PAIRED's Evaluator-R
    subtype; provided it as the anchor for the first worked
    example. \\

DP-E2 & Execution &
    Swimming coaching preprint as the empirical corpus from which
    the three worked examples in Section~4 were drawn. &
    None --- human-originated; the author supplied the
    preprint independently. &
    Selected the preprint as the real-world test case for PAIRED's
    retrospective application on the grounds that its development
    involved substantive and varied human-AI collaboration across
    multiple stages. \\

\hline
\end{tabular}
\end{table}

\begin{table}[h]
\centering
\renewcommand{\arraystretch}{1.6}
\caption{PAIRED micro-log: decision points DP-E3 through DP-R5
from the development of this framework (part 4 of 4). All nineteen
entries across
Tables~\ref{tab:acknowledgments1}--\ref{tab:acknowledgments4}
meet the artifact-trigger criterion --- each artifact appears in
the paper in a form that influenced its final content.}
\label{tab:acknowledgments4}
\begin{tabular}{>{\bfseries}p{1.1cm} >{\bfseries}p{1.8cm} p{3.8cm} p{3.8cm} p{3.8cm}}
\hline
\# & Stage & Artifact adopted & Role label & Human judgment applied \\
\hline

DP-E3 & Execution &
    Retrospective reconstruction of the three decision points from
    memory and documentation, constituting the data acquisition
    process for Section~4's worked examples. &
    Co-ideator --- the author provided the raw recollections;
    the structured reconstruction of each decision point along
    PAIRED's three dimensions emerged collaboratively. &
    Reviewed each reconstructed entry for accuracy against personal
    recollection; approved the characterizations as faithful
    representations of the original interactions. \\

DP-R0 & Reporting &
    Full manuscript drafting: all six sections of the paper plus
    the abstract were drafted by the AI collaborator and reviewed
    by the author. &
    Executor --- the author specified the structure, argument,
    and content requirements at each stage; AI implemented them in
    prose with structural and linguistic judgment. &
    Engaged in three distinct review modes across sections:
    approval, where drafts were accepted as written; critique,
    where specific elements were identified as inaccurate or
    tonally wrong and revised; and augmentation, where the first
    author added substantive content beyond the AI draft. \\

DP-R1 & Reporting &
    Dual format presentation in Section~4: both micro-log table and
    prose rendering shown for each worked example as a demonstration
    of PAIRED's medium-agnostic design principle. &
    Co-ideator --- the strategic reason for showing both formats
    was human-specified; the specific implementation emerged
    collaboratively. &
    Adopted the dual format after recognizing it demonstrated
    PAIRED's flexibility in practice rather than merely asserting
    it in prose. \\

DP-R2 & Reporting &
    Worked example subsection structure: context paragraph,
    micro-log table, prose rendering, and analytical paragraph
    identifying what PAIRED captures that existing frameworks miss. &
    Co-ideator --- the need for a consistent subsection structure
    was human-specified; the four-part sequence emerged
    collaboratively. &
    Adopted the four-part structure after recognizing it moved each
    example from empirical grounding through demonstration to
    analytical payoff in a way that serves both pedagogical and
    argumentative purposes. \\

DP-R3 & Reporting &
    Decision to place the complete PAIRED micro-log in the
    Acknowledgments rather than limiting it to the three
    demonstrative cases from Section~4, establishing the
    Acknowledgments as the first genuine near-prospective
    realization of PAIRED. &
    Co-ideator --- the insufficiency of limiting the log to three
    cases was human-identified; the proposal to use the
    Acknowledgments as the complete prospective record emerged
    collaboratively. &
    Adopted the complete log approach after recognizing that a
    framework paper about honest AI disclosure that selectively
    documented only three of its own decision points would
    undermine its own argument. \\

DP-R4 & Reporting &
    Decision to dedicate a full subsection (Section~5.4) to the
    model-assisted micro-logging proposal, framed explicitly as a call-to-action directed at AI platform developers and
    publishers. &
    Co-ideator --- the author proposed the model-assisted
    logging concept; the decision to elevate it to a standalone
    subsection with a three-audience call-to-action framing emerged
    collaboratively. &
    Adopted the standalone subsection after recognizing that the
    proposal's implications --- for researcher workflow, platform
    architecture, and publisher policy simultaneously. \\


\hline
\end{tabular}
\end{table}

All critical analyses and editorial decisions
in this paper were made by the authors. The authors thoroughly
reviewed all AI-assisted content and take full responsibility for
the integrity and accuracy of the published work.

\bibliographystyle{unsrt}  
\bibliography{references}

\end{document}